\documentclass{Interspeech}
\usepackage{makecell}
\usepackage{xcolor}
\usepackage[subtle]{savetrees}

\interspeechcameraready 
 
\title{BiCrossMamba-ST: Speech Deepfake Detection with Bidirectional Mamba Spectro-Temporal Cross-Attention}

\author[affiliation={1,2}]{Yassine}{El Kheir}
\author[affiliation={1,2}]{Tim}{Polzehl}
\author[affiliation={1,2}]{Sebastian}{Möller}

\affiliation{Speech and Language Technology}{DFKI}{Germany}
\affiliation{Quality and Usability Lab}{Technical University of Berlin}{Germany}
\email{yassine.el\_kheir@dfki.de, tim.polzehl@dfki.de, sebastian.moeller@dfki.de}

\begin{document}

\maketitle

\begin{abstract}

We propose \textbf{BiCrossMamba-ST}, a robust framework for speech deepfake detection that leverages a dual-branch spectro-temporal architecture powered by bidirectional Mamba blocks and mutual cross-attention. By processing spectral sub-bands and temporal intervals separately and then integrating their representations, BiCrossMamba-ST effectively captures the subtle cues of synthetic speech. In addition, our proposed framework leverages a convolution-based 2D attention map to focus on specific spectro-temporal regions, enabling robust deepfake detection. Operating directly on raw features, BiCrossMamba-ST achieves significant performance improvements, a 67.74\% and 26.3\% relative gain over state-of-the-art AASIST on ASVSpoof LA21 and ASVSpoof DF21 benchmarks, respectively, and a 6.80\% improvement over RawBMamba on ASVSpoof DF21. Code and models will be made publicly available.

\end{abstract}

\section{Introduction}

\begin{figure*} 
\centering
\includegraphics[width=0.8\textwidth]{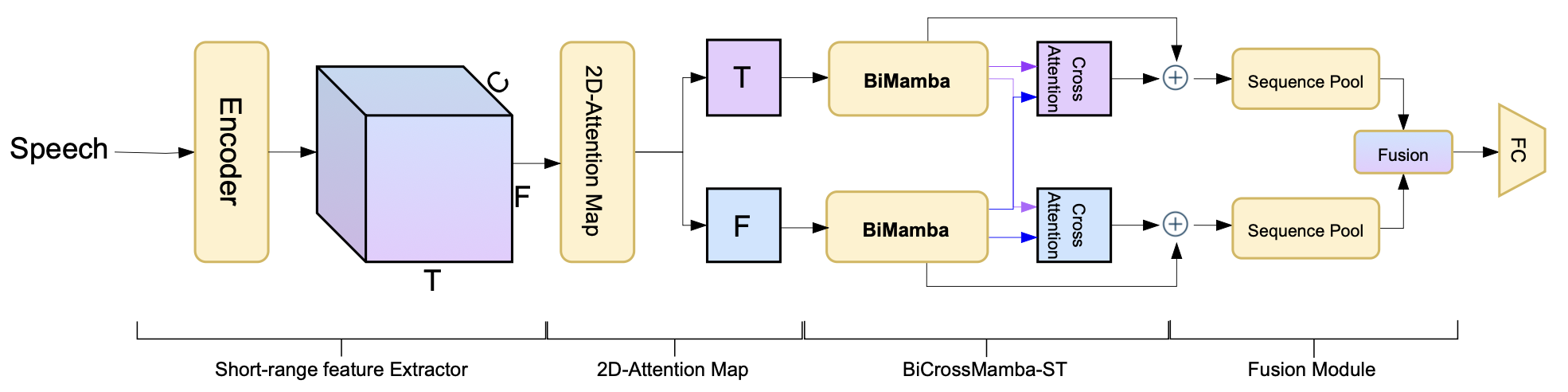}
\vspace{-0.3cm}
\caption{Overview of the proposed BiCrossMamba-ST Framework.}
\label{fig:main_arch}
\vspace{-0.5cm}
\end{figure*}

Recent advancements in speech synthesis and voice conversion technologies have made it increasingly easy to create high-fidelity synthetic audio that convincingly mimics real human voices \cite{kumar2023deep, huang2023singing, yi2023audio}. As a result, distinguishing genuine speech from advanced deepfake audio has become increasingly challenging \cite{khan2022voice}, posing significant risks to Automatic Speaker Verification (ASV) systems that are widely used in critical applications such as access control, telephone banking, and forensic investigations \cite{anjum2017spoofing, li2024audio}.

Research in speech deepfake detection can be broadly categorized into feature-based and end-to-end approaches. Feature-based methods rely on engineered representations—such as the short-time Fourier transform (STFT), linear-frequency cepstral coefficients (LFCC), and constant-Q transform (CQT) \cite{ sahidullah15_interspeech, todisco2017constant, zhang122021effect, li2023investigation, shin2024hm, fan2024dual} to capture distinctive speech characteristics for deepfake detection. In contrast, end-to-end methods optimize directly on raw audio \cite{tak2021graph, jung2022aasist, liu2023leveraging, chen2024rawbmamba}, demonstrating enhanced generalizability and improved performance.

Among end-to-end approaches, a variety of architectures have been proposed to discern the subtle differences between genuine and deepfake speech. For instance, RawNet2 \cite{tak2021end} employs time-domain convolutions to extract discriminative features directly from raw audio signals. However, deepfake artifacts frequently manifest in localized spectro-temporal regions, rendering their detection challenging. To mitigate this limitation, earlier work introduced RawGAT-ST \cite{tak2021end2}, which leverages two parallel graph attention networks (GATs) to concurrently model cues across distinct spectral sub-bands and temporal intervals. The element-wise multiplication of these graph representations yields a unified spectro-temporal representation, enhancing the ability to capture localized artifacts. Despite its efficacy, the inherent heterogeneity between the separately modeled graphs motivated further refinement. ASSIST \cite{jung2022aasist} explored a heterogeneous stacking graph attention layer designed to integrate spectral and temporal graph representations more cohesively. This modification not only reconciles the heterogeneity issue but also results in improved detection performance. On the other hand, a recent work \cite{wu2024spoofing} proposed a mutual cross-attention mechanism as an alternative to the heterogeneous stacking approach, further advancing spectro-temporal modeling with promising results.

Building on these advancements, we explore the potential of Mamba, a state-space-based model \cite{gu2023mamba}, which has recently achieved state-of-the-art performance across various domains, including language modeling \cite{waleffe2024empirical}, computer vision \cite{xu2024survey}, time-series analysis \cite{patro2024simba}, and graph modeling \cite{wang2024graph}. Mamba offers near-linear complexity and a global receptive field, making it a compelling alternative to GAT methods in speech deepfake detection. Unlike GAT networks, which rely on graph pooling strategies to generate more discriminative graphs by selecting a subset of the most informative nodes, Mamba’s selection mechanism enables a more direct and efficient flow of information by dynamically controlling which input features contribute to the hidden states which allow the model to minimize the influence of unimportant nodes at each step of hidden state computation. This implicit node selection facilitates the forwarding of discriminative features crucial for speech deepfake detection while also benefiting reduced memory overhead. Leveraging these strengths, we investigate Mamba’s capability to better localize spectro-temporal artifacts, further enhancing the detection of subtle deepfake cues. To address this, we propose \textbf{BiCrossMamba-ST}, a framework that first applies a 2D-attention map on the feature encoder to help focus on critical spectro-temporal regions. The framework then processes features through two parallel bidirectional Mamba (BiMamba) blocks, modeling spectral and temporal features concurrently. Finally, mutual cross-attention is applied between the two branches to integrate the spectral-branch and temporal-branch features, facilitating effective information exchange between them. The proposed framework is evaluated on four public benchmark databases (i.e., ASVspoof 2019 LA, ASVspoof 2021 LA, ASVspoof 2021 DF, and ASVSpoof5), and results demonstrate the superiority and robustness over SOTA systems. 


\section{Proposed Framework}
\label{sec:framework}
Figure \ref{fig:main_arch} illustrates our proposed framework for speech deepfake detection. The framework processes input audio through four key modules: (1) \textbf{Short-Range Feature Extractor (Encoder)} extracts low-level spectro-temporal features from raw audio through trainable filters, (2) \textbf{2D-Attention Map (2D-AttM)} generates attention maps to highlight critical regions in the feature space, (3) \textbf{BiCrossMamba-ST}, our core dual-branch module that processes spectral and temporal features separately through BiMamba blocks while facilitating information exchange via mutual cross-attention (\textbf{MCA}), and finally (4) \textbf{Fusion Module} integrates the processed features into a unified representation for classification. The following subsections detail each component of our framework.

\begin{table*}[]
\centering
\caption{Experimental results of Mamba derivatives on ASVspoof datasets. Results are the best single system (average over three runs).}
\vspace{-0.3cm}
\scalebox{0.7}{
\begin{tabular}{lccc|ccc|cc|cc}
\toprule

& \multicolumn{3}{@{}c@{}}{\textbf{ASVspoof LA19}} & \multicolumn{3}{@{}c@{}}{\textbf{ASVspoof LA21}} & \multicolumn{2}{@{}c@{}}{\textbf{ASVspoof DF21}}  & \multicolumn{2}{@{}c@{}}{\textbf{ASVspoof5}} \\ \cline{2-11} 
& \textbf{minDCF} & \textbf{EER(\%)} & \textbf{t-DCF} & \textbf{minDCF} & \textbf{EER(\%)} & \textbf{t-DCF} & \textbf{minDCF} & \textbf{EER(\%)} & \textbf{minDCF} & \textbf{EER(\%)} \\ \hline

\textbf{CrossMamba-ST}  & \makecell{0.0341 \\ {\scriptsize (0.0384)}} & \makecell{1.40 \\ {\scriptsize (1.52)}} & \makecell{0.0373 \\ {\scriptsize (0.042)}} & \makecell{0.0943 \\ {\scriptsize (0.0990)}} & \makecell{3.37 \\ {\scriptsize (3.63)}} & \makecell{0.2718 \\ {\scriptsize (0.2772)}} & \makecell{0.4409 \\ {\scriptsize (0.5285)}} & \makecell{16.07 \\ {\scriptsize (18.77)}} & \makecell{0.7454 \\ {\scriptsize (0.7502)}} & \makecell{30.76 \\ {\scriptsize (31.32)}} \\ \hline

\textbf{inBiCrossMamba-ST} & \makecell{0.0354 \\ {\scriptsize (0.0380)}} & \makecell{1.36 \\ {\scriptsize (1.43)}} & \makecell{0.0397 \\ {\scriptsize (0.0429)}} & \makecell{\textbf{0.0848} \\ {\scriptsize (0.1066)}} & \makecell{\textbf{3.30} \\ {\scriptsize (3.97)}} & \makecell{\textbf{0.2618} \\ {\scriptsize (0.2835)}} & \makecell{0.4395 \\ {\scriptsize (0.4687)}} & \makecell{16.26 \\ {\scriptsize (17.76)}} & \makecell{0.6983 \\ {\scriptsize (0.7387)}} & \makecell{29.53 \\ {\scriptsize (31.25)}} \\ \hline

\textbf{FlipCrossMamba-ST} & \makecell{0.0319 \\ {\scriptsize (0.0382)}} & \makecell{1.33 \\ {\scriptsize (1.54)}} & \makecell{0.0358 \\ {\scriptsize (0.0425)}} & \makecell{0.0876 \\ {\textbf{\scriptsize (0.0946)}}} & \makecell{3.34 \\ {\textbf{\scriptsize (3.59)}}} & \makecell{0.2653 \\ {\textbf{\scriptsize (0.2715)}}} & \makecell{0.4455 \\ {\scriptsize (0.4808)}} & \makecell{17.84 \\ {\scriptsize (18.56)}} & \makecell{0.6994 \\ {\scriptsize (0.7000)}} & \makecell{\textbf{29.14} \\ {\scriptsize (29.16)}} \\ \hline

\textbf{BiCrossMamba-ST} & \makecell{\textbf{0.0281} \\ {\scriptsize (0.0331)}} & \makecell{\textbf{1.08} \\ {\scriptsize (1.26)}} & \makecell{\textbf{0.0312} \\ {\scriptsize (0.0372)}} & \makecell{0.0870 \\ {\scriptsize (0.1073)}} & \makecell{3.39 \\ {\scriptsize (3.99)}} & \makecell{0.2639 \\ {\scriptsize (0.2841)}} & \makecell{\textbf{0.3942} \\ {\scriptsize (0.4401)}} & \makecell{\textbf{14.77} \\ {\scriptsize (16.20)}} & \makecell{\textbf{0.6884} \\ {\scriptsize (0.6887)}} & \makecell{29.53 \\ {\scriptsize (29.84)}} \\ \bottomrule

\end{tabular}}
\label{tab:variations_mamba}
\end{table*}

\begin{table*}[]
\centering
\vspace{-0.2cm}
\small\caption{Ablation study results for BiCrossMamba-ST on ASVspoof datasets. Results are best single system (average over three runs).}
\vspace{-0.3cm}
\scalebox{0.7}{
\begin{tabular}{lccc|ccc|cc}
\toprule
& \multicolumn{3}{c}{\textbf{ASVspoof LA19}} & \multicolumn{3}{c}{\textbf{ASVspoof LA21}} & \multicolumn{2}{c}{\textbf{ASVspoof DF21}} \\ \cline{2-9}
& \textbf{minDCF} & \textbf{EER(\%)} & \textbf{tDCF} & \textbf{minDCF} & \textbf{EER(\%)} & \textbf{tDCF} & \textbf{minDCF} & \textbf{EER(\%)} \\ \hline

\textbf{BiCrossMamba-ST} & 
\makecell{\textbf{0.0281} \\ {\scriptsize (0.0331)}} & 
\makecell{\textbf{1.08} \\ {\scriptsize (1.26)}} & 
\makecell{\textbf{0.0312} \\ {\scriptsize (0.0372)}} & 
\makecell{\textbf{0.0870} \\ {\scriptsize (0.1073)}} & 
\makecell{\textbf{3.39} \\ {\scriptsize (3.99)}} & 
\makecell{\textbf{0.2639} \\ {\scriptsize (0.2841)}} & 
\makecell{\textbf{0.3942} \\ {\scriptsize (0.4401)}} & 
\makecell{\textbf{14.77} \\ {\scriptsize (16.20)}} \\ \hline

\textbf{}                 & \multicolumn{8}{c}{\textbf{Ablations 1: BiMamba vs GAT vs Transformer}}  \\ \hline

\textbf{w Transformer} & 
\makecell{0.0442 \\ {\scriptsize (0.0452)}} & 
\makecell{1.76 \\ {\scriptsize (1.75)}} & 
\makecell{0.0491 \\ {\scriptsize (0.0502)}} & 
\makecell{0.1216 \\ {\scriptsize (0.1284)}} & 
\makecell{4.42 \\ {\scriptsize (4.71)}} & 
\makecell{0.2981 \\ {\scriptsize (0.3040)}} & 
\makecell{0.4459 \\ {\scriptsize (0.4533)}} & 
\makecell{16.86 \\ {\scriptsize (17.45)}} \\ 
\cline{2-9}
\textbf{w GAT} & 
\makecell{0.0286 \\ {\scriptsize (0.0352)}} & 
\makecell{1.29 \\ {\scriptsize (1.45)}} & 
\makecell{0.0385 \\ {\scriptsize (0.0389)}} & 
\makecell{0.1316 \\ {\scriptsize (0.1399)}} & 
\makecell{4.65 \\ {\scriptsize (4.93)}} & 
\makecell{0.3092 \\ {\scriptsize (0.3173)}} & 
\makecell{0.5585 \\ {\scriptsize (0.6557)}} & 
\makecell{19.54 \\ {\scriptsize (23.16)}} \\ \hline

\textbf{}                 & \multicolumn{8}{c}{\textbf{Ablations 2: BiCrossMamba-ST Modules}} \\ \hline

\textbf{w/o 2D-AttM} & 
\makecell{0.0320 \\ {\scriptsize (0.0371)}} & 
\makecell{1.36 \\ {\scriptsize (1.46)}} & 
\makecell{0.0350 \\ {\scriptsize (0.0413)}} & 
\makecell{0.1341 \\ {\scriptsize (0.1578)}} & 
\makecell{4.68 \\ {\scriptsize (5.58)}} & 
\makecell{0.3123 \\ {\scriptsize (0.3359)}} & 
\makecell{0.4421 \\ {\scriptsize (0.5087)}} & 
\makecell{16.83 \\ {\scriptsize (18.47)}} \\ 
\cline{2-9}
\textbf{w/o MCA} & 
\makecell{0.0338 \\ {\scriptsize (0.0369)}} & 
\makecell{1.32 \\ {\scriptsize (1.47)}} & 
\makecell{0.0376 \\ {\scriptsize (0.0407)}} & 
\makecell{0.0905 \\ {\scriptsize (0.1057)}} & 
\makecell{3.77 \\ {\scriptsize (4.07)}} & 
\makecell{0.2656 \\ {\scriptsize (0.2813)}} & 
\makecell{0.4071 \\ {\scriptsize (0.4276)}} & 
\makecell{15.82 \\ {\scriptsize (16.90)}} \\ 
\cline{2-9}
\textbf{w/o spectral} & 
\makecell{0.0335 \\ {\scriptsize (0.0395)}} & 
\makecell{1.28 \\ {\scriptsize (1.57)}} & 
\makecell{0.0367 \\ {\scriptsize (0.0429)}} & 
\makecell{0.1200 \\ {\scriptsize (0.1864)}} & 
\makecell{4.37 \\ {\scriptsize (6.57)}} & 
\makecell{0.2984 \\ {\scriptsize (0.3646)}} & 
\makecell{0.5741 \\ {\scriptsize (0.582)}} & 
\makecell{20.59 \\ {\scriptsize (21.22)}} \\ 
\cline{2-9}
\textbf{w/o temporal} & 
\makecell{0.0368 \\ {\scriptsize (0.042)}} & 
\makecell{1.40 \\ {\scriptsize (1.60)}} & 
\makecell{0.0418 \\ {\scriptsize (0.0470)}} & 
\makecell{0.0930 \\ {\scriptsize (0.1251)}} & 
\makecell{3.47 \\ {\scriptsize (4.50)}} & 
\makecell{0.2691 \\ {\scriptsize (0.3017)}} & 
\makecell{0.4966 \\ {\scriptsize (0.5337)}} & 
\makecell{18.03 \\ {\scriptsize (19.69)}} \\ 
 \bottomrule

\end{tabular}}
\label{tab:mamba_performance}
\vspace{-0.4cm}
\end{table*}

\subsection{Encoder}
\label{encoder}

For the feature encoder, we adopt a modified version of RawNet2 \cite{liu2023leveraging} to extract rich spectro-temporal representations from raw audio. The encoder first employs parametric sinc functions as learnable band-pass filters (70 filters) to transform the raw waveform into low-level feature maps (LFM) $\in \mathbb{R}^{f \times t}$, where $f$ and $t$ represent frequency and temporal bins respectively. These features are then enhanced through a series of convolution modules: first processed by a ResNet block with $32$ filters, followed by three ResSERes2Net blocks \cite{li2021replay} with $64$ filters. The ResSERes2Net blocks incorporate squeeze-and-excitation operations to adaptively enhance channel-wise feature representations, producing refined high-level feature maps $x_{HFM} \in \mathbb{R}^{C \times F \times T}$. Here, $C$, $F$, and $T$ denote the number of channels, frequency bins, and temporal bins respectively.

\subsection{2D-AttM Module}  

To enhance the processing of high-level feature maps \( x_{HFM} \in \mathbb{R}^{C \times F \times T} \), we use a learnable \textbf{2D attention mechanism} that dynamically identifies and emphasizes critical spectro-temporal regions \cite{wu2024spoofing}. Unlike conventional methods that rely on average or max pooling \cite{jung2022aasist}, this approach generates adaptive attention maps that assign higher weights to the relevant regions of the presentation $x_{HFM}$.  

The attention map \( M \) is computed using a two-step convolutional process:  
\[
M = \text{Softmax}\left( \text{Conv2D}\left( \text{SeLU}\left( \text{Conv2D}(x_{HFM}) \right) \right) \right)
\]

This produces a mask \( M \) of size \( F \times T \), capturing the relative significance of different regions within the feature space. The mask is then applied to \( x_{HFM} \) (e.g. element-wise multiplication), followed by domain-specific summation to generate two complementary feature representations:  
\textbf{Spectral branch} \( x_F \in \mathbb{R}^{F \times C} \) and \textbf{Temporal branch} \( x_T \in \mathbb{R}^{T \times C} \).

\begin{figure} 
\centering
\includegraphics[width=0.45\textwidth]{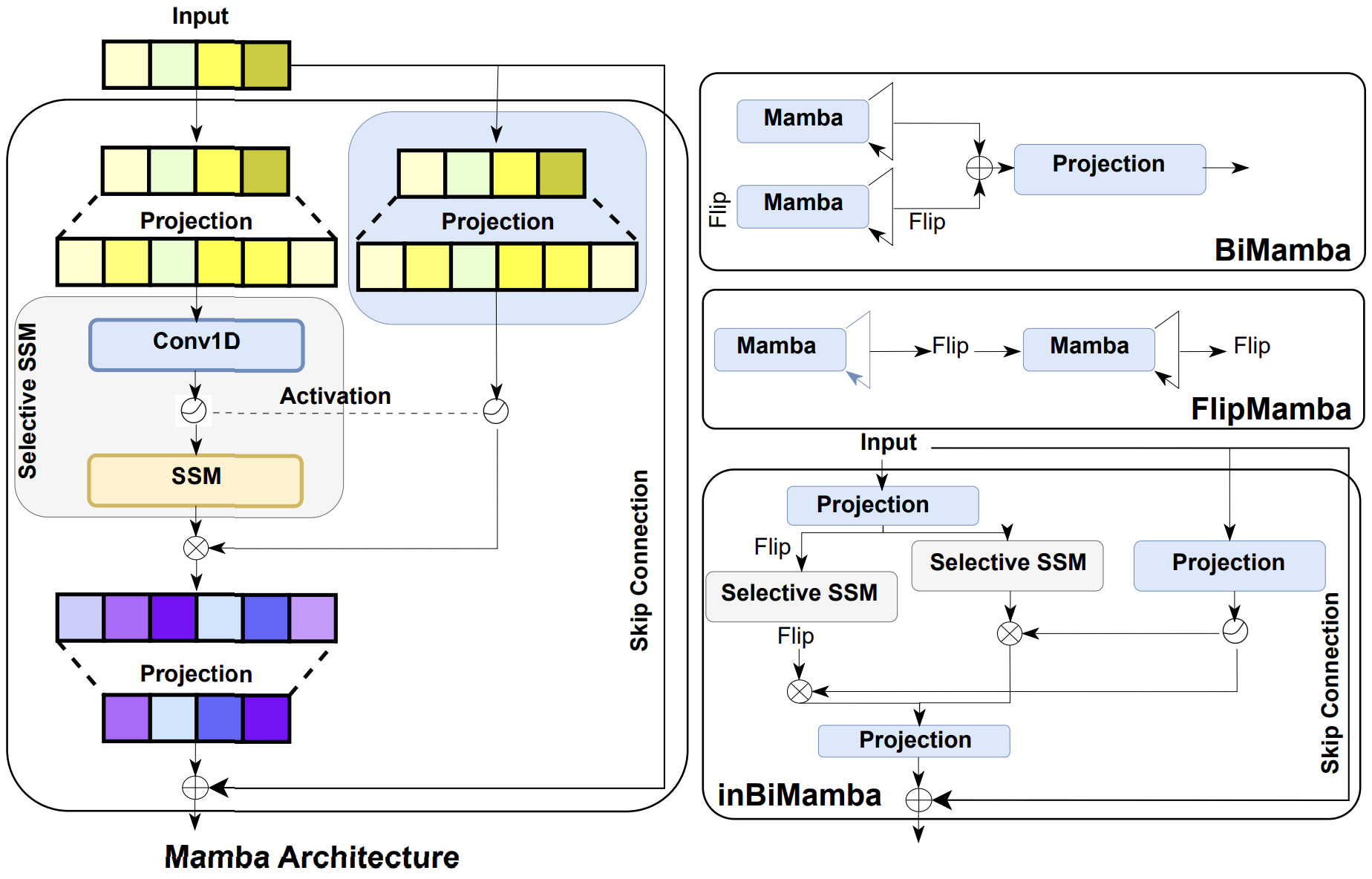}
\vspace{-0.2cm}
\caption{Mamba Architecture and Its Derivatives: \textbf{SSM} refers to Selective State Space Models, while \textbf{Flip} denotes the reversal of input sequences.}
\label{fig:mamba_ens}
\vspace{-0.5cm}
\end{figure}

\subsection{BiCrossMamba-ST Module}
\label{xmamba}

BiCrossMamba-ST, illustrated in Figure~\ref{fig:main_arch}, is a dual-branch spectro-temporal module that employs mutual cross-attention (MCA) to enhance feature interactions. The spectral ($x_F$) and temporal ($x_T$) features are initially processed independently using the BiMamba blocks. MCA mechanism then facilitates the exchange of relevant information between the resultant $x_F$ and $x_T$. Below, we provide a detailed explanation of the BiMamba block and the MCA module.

\subsubsection{Mamba, BiMamba, and other derivatives}

State Space Models (SSMs), akin to Recurrent Neural Networks (RNNs), map input sequences \( x(t) \in \mathbb{R}^N \) to output sequences \( y(t) \in \mathbb{R}^N \) via a hidden state \( h(t) \in \mathbb{R}^N \). This process is controlled by a linear ordinary differential equation (ODE):

\scalebox{0.9}{%
\begin{minipage}{\linewidth}
\[
h'(t) = A\,h(t) + B\,x(t),\quad y(t) = C\,h(t).
\]
\end{minipage}
}
\vspace{0.2cm}


where \( A \), \( B \), and \( C \) are the state, input, and output matrices, respectively. Here, \( A h(t) \) models state evolution, \( B x(t) \) captures input influence, and \( C h(t) \) translates the state into output. Discretization involves adjusting matrices \( A \) and \( B \) with step size \( \Delta \):
\scalebox{0.9}{
\begin{minipage}{\linewidth}
\begin{equation*}
A = \exp(\Delta A), \quad B = (\Delta A)^{-1} (\exp(\Delta A) - I) \Delta B.
\end{equation*}
\end{minipage}
}

\vspace{0.2cm}

Although SSMs and their enhanced variant, Structured State Space Sequence (S4), excel at representing long sequences, their static nature limits adaptability to varying input contexts. To address this, Mamba introduces parameterization of matrices \( B \), \( C \), and \( \Delta \), enabling dynamic adjustment based on input \( x_t \). This allows, \( B \) to control over input influence on the hidden state, \( C \) to control over how the hidden state affects the output, and \( \Delta \): to control the degree of emphasis on current inputs versus past states. However, SSMs' unidirectional processing restricts global contextual modeling. To overcome this limitation, we propose three bidirectional derivatives: BiMamba, FlipMamba, and inBiMamba (see Figure~\ref{fig:mamba_ens}).

\vspace{-0.3cm}
\paragraph*{\textbf{BiMamba:}}
\label{bimamba}

As shown in Figure \ref{fig:mamba_ens}, BiMamba~\cite{zhang2024mamba} utilizes two parallel Mamba blocks: one processes the input as-is while the other processes the flipped (reversed) sequence. The outputs are combined by reversing back the second block's output, concatenating both representations and projecting them into a unified representation. Our proposed \textbf{BiCrossMamba-ST} framework uses BiMamba. The corresponding framework that uses the original Mamba is named \textbf{CrossMamba-ST}.

\vspace{-0.4cm}

\paragraph*{\textbf{FlipMamba:}}

FlipMamba as shown in Figure \ref{fig:mamba_ens}, processes inputs sequentially through two Mamba blocks. The first block processes the sequence; its output is flipped and passed into the second Mamba block. The final output is flipped back to its original order. This sequential flipping mechanism ensures richer bidirectional information flow. The corresponding framework that uses FlipMamba is named \textbf{FlipCrossMamba-ST}.
\vspace{-0.4cm}

\paragraph*{\textbf{inBiMamba:}}
Inspired by~\cite{zhu2024vision}, inBiMamba employs two Selective SSM modules sharing common input and output projection layers. One SSM module processes the input sequence directly while the other processes the flipped version. Representations are then combined and then passed to a projection layer to form a unified representation. The corresponding framework that uses inBiMamba is named \textbf{inBiCrossMamba-ST}.

\subsubsection{MCA Module} 
After independently processing the spectral $x_F$ and temporal $x_T$ branches using BiMamba (or other proposed derivatives), we introduce the MCA module mechanism to facilitate effective spectro-temporal information exchange. We used single-head attention \cite{vaswani2017attention}, this mechanism enables cross-branch feature interaction through adaptive feature recalibration. Each branch-spectral or temporal-can selectively emphasize or suppress parts of the other branch's representation based on their mutual relevance. Specifically, the spectral branch uses the temporal branch's embeddings as keys and values, and vice versa.
\scalebox{0.9}{%
\begin{minipage}{\linewidth}
\begin{align*}
x_F &= \text{LayerNorm}(x_F + \text{Attention}(Q=x_F, K=x_T, V=x_T)) \\
x_T &= \text{LayerNorm}(x_T + \text{Attention}(Q=x_T, K=x_F, V=x_F))
\end{align*}
\end{minipage}%
}

\subsection{Fusion Module} After MCA Module, we apply sequence pooling \cite{li2022role} to extract key information from resultant spectral (\(x_F\)) and temporal (\(x_T\)) features:




\vspace{-0.1cm}

\scalebox{0.9}{%
\hspace{-0.5cm}
\begin{minipage}{\linewidth}
\begin{align*}
z_F &= \text{Softmax}(g_F(x_F))\cdot x_F \in \mathbb{R}^{C}, \quad
z_T=\text{Softmax}(g_T(x_T))\cdot x_T \in \mathbb{R}^{C}.
\end{align*}
\end{minipage}
}


\noindent where $g$ is a linear layer with one unit, and $z_F$ and $z_T$ are respective weighted averages. We concatenate these vectors and apply a linear projection to obtain a final representation of dimension (\#C).

\section{Experimental Setup}
\label{sec_exp}

\begin{table*}[]
\centering
\caption{Comparison with SOTA single systems. *:results from \cite{chen2024rawbmamba}. \(^\dagger\) Reproduced results. The minDCT metric is not reported for ASVSpoof LA19, LA21, and DF21 because SOTA models used for comparison do not include it (newly used metric).}
\vspace{-0.3cm}
\scalebox{0.75}{
\begin{tabular}{lcc|cc|c|cc}
\toprule              

&  \multicolumn{2}{c}{\textbf{ASVspoof LA19}}   & \multicolumn{2}{c}{\textbf{ASVspoof LA21}}   & \multicolumn{1}{c}{\textbf{ASVspoof DF21}}  & \multicolumn{2}{c}{\textbf{ASVspoof5}}  \\ \cline{2-8} 
               & \textbf{EER(\%)} & \textbf{t-DCF} & \textbf{EER(\%)} & \textbf{t-DCF}  & \textbf{EER(\%)} & \textbf{minDCF} & \textbf{EER(\%)} \\ \hline

RawNet2 \cite{tak2021end}            & 1.14       & -       & 9.50             & -                & 22.38 & 0.8266     & 36.04  \\
AASIST\(^\dagger\) \cite{jung2022aasist}           & \textbf{0.93}    & \textbf{0.0285}  & 10.51            & 0.4884           & 20.04   & 0.7106    & \textbf{29.12}       \\
AASIST-4Block \cite{jung2022aasist}      & 1.20              & 0.0341           & 9.15             & 0.437            & -    & -    & -             \\
ARawNet2 \cite{li2023advanced}            & 4.61       & -       & 8.36             & -                & 19.03  & -     & -   \\

SE-Rawformer\(^\dagger\) \cite{liu2023leveraging}       & 1.15             & 0.0314           & 4.31             & 0.2851           & 20.26     & -     & -        \\
HM-Conformer \cite{shin2024hm}       & -           & -           & -            & -           & 15.71      & -     & -     \\
RawBMamba  \cite{chen2024rawbmamba}       & 1.19             & 0.0360            & \textbf{3.28}    & 0.2709  & 15.85     & 0.7120\(^\dagger\)     & 30.77\(^\dagger\)     \\ \hline
\textbf{BiCrossMamba-ST} (ours) & 1.08             & 0.0312           & 3.39             & \textbf{0.2639}  & \textbf{14.77} & \textbf{0.6884}     & 29.53  \\ \bottomrule 
\end{tabular}}
\vspace{-0.5cm}
\label{tab:comparaison_ours}
\end{table*}

\subsection{Dataset and evaluation metrics}
Our experiments are conducted on the ASVspoof LA19 \cite{wang2020asvspoof}, ASVspoof LA21, ASVspoof DF21 \cite{yamagishi2021asvspoof}, and ASVspoof5 \cite{wang2024asvspoof5} datasets to evaluate our model's performance and generalization. We train our primary models using the ASVspoof LA19 training set, a widely adopted benchmark in speech anti-spoofing research. We then evaluate these models on the ASVspoof LA19, ASVspoof LA21, and ASVspoof DF21 datasets. ASVspoof LA21 dataset reflects real-world telephony conditions with various speech codecs, while the ASVspoof DF21 dataset consists of data encoded with lossy codecs intended for media storage, allowing us to assess generalization in realistic scenarios. In addition, to further analyze performance on challenging deepfake detection tasks, we report results training and evaluating the ASVspoof5 dataset, which contains stronger spoofing attacks optimized to fool models through adversarial techniques. Our evaluation employs standard metrics, including the minimum detection cost function (minDCT) \cite{wang2024asvspoof5} and the Equal Error Rate (EER), with the minimum tandem detection cost function (tDCT) \cite{kinnunen2020tandem} also reported for LA19 and LA21 to provide a comprehensive assessment.

\subsection{Implementation Details}
During training, the input data comprises 64,600 sampling points ($\approx$ 4 seconds). All models are trained for 300 epochs using Adam optimizer \cite{diederik2014adam} with a batch size of 32, an initial learning rate of \(5 \times 10^{-4}\), and a weight decay of \(1 \times 10^{-4}\). The ASoftmax loss function \cite{liu2017sphereface} is employed, and training is conducted on the same single H100-GPU for all experiments. To ensure reliable results, all experiments are conducted three times with different random seeds. We report the best single system along with the average results over three runs.

\section{Experimental results}
\label{sec_res}

\subsection{Our framework with different Mamba derivatives}

In Section~\ref{xmamba}, we introduced several Mamba derivatives for processing spectral and temporal features within our framework. In this section, we compare these four variants: \textbf{CrossMamba-ST} (using Mamba), \textbf{inBiCrossMamba-ST} (using inBiMamba), \textbf{FlipCrossMamba-ST} (using FlipMamba), and \textbf{BiCrossMamba-ST} (using BiMamba). Table~\ref{tab:variations_mamba} reports the performance of these models.
The CrossMamba-ST already delivers competitive results on in-domain ASVspoof LA19 data. However, incorporating bidirectional processing brings notable benefits. In particular, the BiCrossMamba-ST model achieves the best performance on ASVspoof LA19, showing a relative improvement of 17.6\% in minDCF over CrossMamba-ST. On the ASVspoof LA21 dataset—which reflects real-world telephony conditions—all bidirectional variants perform similarly, with inBiCrossMamba-ST showing a slight edge. More importantly, on the ASVspoof DF21 dataset—which reflects real-world media storage conditions—, BiCrossMamba-ST markedly outperforms the baseline and other models by reducing both minDCF and EER by at least 10.5\% and 8\%, respectively, demonstrating its efficient generalization capabilities. For the challenging ASVspoof5 dataset, although EERs remain high across all models, the bidirectional derivatives exhibit substantial improvements in minDCF compared to CrossMamba-ST, with BiCrossMamba-ST attaining the best minDCT of $0.6884$, a 7.6\% relative improvement. Based on these insights, the BiCrossMamba-ST framework—with its bidirectional Mamba BiMamba design—provides the best balance between in-domain performance and generalization capabilities\footnote{Increasing the number of Mamba blocks in our proposed frameworks does not yield further improvements.}. 

\subsection{BiMamba vs GAT vs Transformer}

To demonstrate the effectiveness of BiMamba in processing spectral and temporal features for speech deepfake detection, we replaced the BiMamba block in our main framework (see Figure~\ref{fig:main_arch}) with two popular alternatives: a GAT network following the design in \cite{wu2024spoofing} and Transformers module as used in \cite{liu2023leveraging}. The remainder of the framework is identical across all configurations. As detailed in Table \ref{tab:mamba_performance}, replacing BiMamba with Transformers (w Transformer) leads to a relative degradation in performance, with the ASVspoof LA19 relative minDCF increasing by 57.3\%, and the ASVspoof DF21 relative EER rising by 14.1\%. Similarly, the GAT (w GAT) model shows even larger relative performance drops, with ASVspoof DF21 relative minDCF rising by 41.7\% and EER by 32.3\%. These results underscore the advantage of BiMamba, as its selection mechanism captures discriminative spectral-temporal features, enhancing speech deepfake detection and generalization.

\subsection{Ablation study of BiCrossMamba-ST modules}

We conducted an ablation study to assess the contributions of BiCrossMamba-ST framework key modules: 2D-AttM, MCA, and its spectral and temporal branches. As shown in Table \ref{tab:mamba_performance} Ablation 2, removing the 2D-AttM (w/o 2D-AttM) module led to a relative increase of 12.15\% in minDCF and 14.0\% in EER on ASVspoof DF21, underscoring its critical role in focusing on discriminative spectro-temporal regions. Ablating the MCA module (w/o MCA) resulted in less severe degradations but still notable impact, with ASVspoof LA21 relative minDCF rising by 4.0\% and ASVspoof DF21 EER by 7.1\%, confirming its value in refining cross-domain representations. The spectral branch proved particularly more important; its removal (w/o spectral) caused a severe performance drop, with ASVspoof DF21 relative minDCF increasing by 45.6\% and EER by 39.4\%. In contrast, eliminating the temporal branch led to a less pronounced degradation, with ASVspoofDF21 relative minDCF and EER rising by 26.0\% and 22.1\%, respectively. These results highlight the greater importance of the spectral branch compared to the temporal branch, but that both are beneficial, aligning with previous findings \cite{tak2021end2, wu2024spoofing}.


\subsection{Comparison with other end-to-end models}

Comparative experiments as shown in Table \ref{tab:comparaison_ours} demonstrate the BiCrossMamba-ST framework's robust performance compared to other SOTA models. On ASVSpoof LA19, Our proposed achieves near-SOTA performance, with an EER of $1.08\%$. Notably, significant improvements are observed in real-world scenarios, on ASVSpoof LA21 and DF21, we show relative EER reductions of $67.74\%$ and $26.3\%$ over AASIST, respectively. 

When compared with alternative modeling techniques like SE-Rawformer and RawBMamba—which flatten RawNet features into a two-dimensional representation $x_{HFM} \in \mathbb{R}^{C \times FT}$ processed via Transformer or bidirectional Mamba blocks—BiCrossMamba-ST consistently outperforms or reaches comparable results to these approaches. On ASVSpoof LA21, it achieves a $21.3\%$ relative improvement over SE-Rawformer and the lowest t-DCF. Notably for ASVSpoof DF21, we achieved better generalization surpassing previous approaches by $6\%$ to $27\%$ relative EER improvement, including the HM-Conformer method that uses additional data augmentations techniques.


The challenging ASVspoof5 benchmark further validates our framework's robustness. While all systems yield EERs above $29\%$, BiCrossMamba-ST achieves the best minDCF, outperforming RawBMamba and AASIST. These results are achieved using only $4$ Mamba blocks\footnote{Two BiMamba blocks, each consisting of two Mamba blocks} and a total of $516K$ parameters, making our BiCrossMamba-ST framework $28.2\%$ lighter than RawBMamba, which uses $12$ blocks and $719K$ parameters.

\section{Conclusion}
\label{conc}


In this paper, we introduce BiCrossMamba-ST, a novel speech deepfake detection framework leveraging BiMamba architecture with cross-attention mechanisms to effectively exploit spectral and temporal features. Our approach outperforms existing Transformer-based and GAT models, demonstrating superior performance across challenging datasets, particularly ASVspoof DF21. The ablation study validates the critical contributions of 2DAttM, and MCA modules in enhancing detection capabilities. For future work, we aim to further explore the effectiveness of BiCrossMamba-ST with speech self-supervised learning features to improve robustness over ASVSpoof5, develop more efficient mutual cross-attention mechanisms using Mamba instead of multi-head attention, and further enhance the 2D-attention maps.

\bibliographystyle{IEEEtran}
\bibliography{mybib}

\end{document}